\documentclass[12pt]{article}

\usepackage{epsfig}
\usepackage{amssymb,amsmath}
\usepackage{multirow}
\usepackage{rotating}
\usepackage{graphicx}
\usepackage{lscape}
\usepackage{hyperref}

\newcommand{\bea}{\begin{eqnarray}}
\newcommand{\eea}{\end{eqnarray}}

 \makeatletter
\def\alt{\mathrel{\mathpalette\gl@align<}}
\def\agt{\mathrel{\mathpalette\gl@align>}}
\def\gl@align#1#2{\lower.6ex\vbox{\baselineskip\z@skip\lineskip\z@
\ialign{$\m@th#1\hfil##\hfil$\crcr#2\crcr\sim\crcr}}} \makeatother

\begin{document}

\begin{flushright}
%preprint number \\
\end{flushright}

\vspace*{1.0cm}

\begin{center}
\baselineskip 20pt 
{\Large\bf 
Positively deflected anomaly mediation 
in the light of the Higgs boson discovery
}
\vspace{1cm}

{\large 
Nobuchika Okada$^{a,}$\footnote{ E-mail: okadan@ua.edu} 
and 
Hieu Minh Tran$^{b,}$\footnote{ E-mail: hieu.tranminh@hust.edu.vn} 
} 
\vspace{.5cm}

{\baselineskip 20pt \it
$^a$Department of Physics and Astronomy, University of Alabama, \\ 
Tuscaloosa, Alabama 35487, USA \\
\vspace{2mm} 
$^b$Hanoi University of Science and Technology, 
1 Dai Co Viet Road, Hanoi, Vietnam \\
}

\vspace{.5cm}

\vspace{1.5cm} {\bf Abstract}
\end{center}

Anomaly-mediated supersymmetry breaking (AMSB) 
 is a well-known mechanism for flavor-blind transmission 
 of supersymmetry breaking from the hidden sector 
 to the visible sector. 
However, the pure AMSB scenario suffers from a serious drawback, 
 namely, the tachyonic slepton problem, 
 and needs to be extended. 
The so-called (positively) deflected AMSB is a simple extension 
 to solve the problem and also provides us with the usual 
 neutralino lightest superpartner 
 as a good candidate for dark matter in the Universe. 
Motivated by the recent discovery of the Higgs boson 
 at the Large Hadron Collider (LHC) experiments, 
 we perform the parameter scan in the deflected AMSB scenario 
 by taking into account a variety of phenomenological constraints 
 such as the dark matter relic density and the observed Higgs boson mass 
 around 125-126 GeV. 
We identify the allowed parameter region and 
 list benchmark mass spectra. 
We find that in most of the allowed parameter regions, 
 the dark matter neutralino is Higgsino-like 
 and its elastic scattering cross section 
 with nuclei is within the future reach of 
 the direct dark matter search experiments, 
 while (colored) sparticles are quite heavy and their discovery 
 at the LHC is challenging.

\thispagestyle{empty}

%\bigskip
\newpage

\addtocounter{page}{-1}

%%%%%%%%%%%%%%%%%%%%%%%%%%
%\baselineskip 36pt
% Main body
%%%%%%%%%%%%%%%%%%%%%%%%%%
\baselineskip 18pt
%%%%%%%%%%%%%%%%%%%%%%%%%%

%%%%%%%%%%%%%%%%%%%%%%%%%%
\section{Introduction}  
%%%%%%%%%%%%%%%%%%%%%%%%%%

%- New LHC result
%- Higgs and SUSY

Recently, the ATLAS \cite{atlas} and CMS \cite{cms} collaborations 
 announced the discovery of a new scalar particle 
 at the Large Hadron Collider (LHC), 
 which is most likely the Standard Model (SM) Higgs boson, 
 with a mass measured, respectively, as 
\begin{eqnarray}
m_h &=& 126.0 \pm 0.4(\text{stat.}) \pm 0.4(\text{syst.}) 
 \, \text{GeV} , \label{ATLAShiggs} \\ 
m_h &=& 125.3 \pm 0.4(\text{stat.}) \pm 0.5(\text{syst.}) 
 \, \text{GeV}. 
\label{CMShiggs}  
\end{eqnarray}
Although we need more data accumulation to conclude that
 it is truly the SM Higgs boson, these observations 
 have ignited a new trend of particle physics research. 
Since the first announcement from CERN about the signal excess 
 in the Higgs boson searches at the LHC, 
 the implication of a 125 GeV Higgs boson has been intensively studied 
 for about a year, in particular, the implications for supersymmetric
 (SUSY) models \cite{ImplicationHiggs}.

SUSY extension of the SM is one of the most promising ways 
 to solve the gauge hierarchy problem, 
 because the quadratic divergences of the Higgs self-energy corrections 
 are completely canceled out by contributions between superpartners 
 and hence SUSY models are insensitive to ultraviolet (UV) physics. 
Moreover in the minimal supersymmetric extension of the SM (MSSM) 
 with R-parity conservation, the lightest superpartner (LSP), 
 usually the neutralino, is a weakly interacting massive particle 
 and a good candidate for dark matter in the Universe.

%- SUSY breaking

%Requirement from experiment, SUSY breaking, origin of soft term, 

Since none of experiments has directly found superpartners, 
 SUSY must be broken at low energies. 
In addition, many results of the indirect search for superpartners 
 require a very special way of generating the soft SUSY-breaking terms, namely, they must be almost flavor-blind 
 and CP-invariant. 
There are several mechanisms that can generate 
 such soft SUSY breaking terms naturally, such as 
 the gauge-mediated SUSY breaking (GMSB) \cite{GMSB} 
 and anomaly-mediated SUSY breaking (AMSB) \cite{AMSB1, AMSB2}.

%What we focus on in this paper: AMSB

In this paper, we focus on the AMSB scenario 
 where SUSY breaking is mediated to the MSSM sector 
 through the superconformal anomaly. 
In this scenario, the nonzero vacuum expectation value (VEV) 
 of the $F$ component of the compensating multiplet ($F_\phi$)
 is the unique SUSY-breaking source that results in 
 all types of soft terms. 
This scenario is based on supergravity and 
 hence AMSB contributions to soft terms always exist. 
In the usual supergravity scenario,
 $F_\phi$ coincides with the gravitino mass, and  
 the gravitino is much heavier than other sparticles 
 in the AMSB scenario
 (see Ref.~\cite{almostNS} for an exception). 
Unfortunately, the pure AMSB scenario cannot be realistic 
 because the slepton squared masses are predicted to be negative. 
There are several proposed solutions to this tachyonic slepton problem
 in the AMSB scenario with simple modifications  
 from the pure AMSB \cite{negativeAMSB,KSS,Jack:2000cd,positiveAMSB,Sundrum:2004un}. 

%- Deflected AMSB, previous research

We consider the scenario involving the so-called deflected anomaly 
 mediation \cite{negativeAMSB, positiveAMSB}. 
The basic idea is to introduce a messenger sector 
 as in the GMSB, and the threshold corrections 
 by the messenger fields deflect the renormalization 
 group (RG) trajectory of the soft terms from those 
 in the pure AMSB scenario. 
Although this scenario looks similar to the GMSB, 
 there is a crucial difference, namely, the SUSY breaking 
 in the messenger sector is generated by the AMSB 
 and hence the scenario is basically the AMSB scenario. 
Thus, the SUSY breaking generated in the messenger sector 
 is proportional to the original SUSY-breaking source: 
 the $F$-term of the compensating multiplet. 
The constant of this proportionality 
 is called the ``deflection parameter'', introduced in Ref. \cite{positiveAMSB}, 
 which is a measure of how much the model is deflected 
 from the pure AMSB. 
Since the SUSY breaking in the messenger  sector is secondary, 
 the deflection parameter should be of $\mathcal{O}(1)$, at most, 
 for the sake of the theoretical consistency. 
The deflection parameter can be either negative \cite{negativeAMSB} 
 or positive \cite{positiveAMSB}. 
Both the resultant sparticle mass spectrum and the 
 cosmological aspects of the scenario 
 are quite different between these two cases. 
While the LSP was found to be a new particle 
 for a negative deflection parameter 
 \cite{negativeAMSB, negativeAMSB-pheno}, 
 the deflected AMSB with a positive deflection parameter
 provides us with the lightest neutralino as the LSP, 
 as usual in the MSSM \cite{positiveAMSB}. 
Thus, in this paper we consider the positively 
 deflected AMSB scenario.

The phenomenology of the positively deflected AMSB scenario, 
 especially the dark matter physics has been previously 
 investigated in Ref.~\cite{neutralinoDM}. 
However, very large values of the deflection parameter $ \gg 1$ 
 were taken into account in the previous analysis\footnote{
In this case, the particle mass spectrum is similar to 
 the one in the GMSB, but with heavy gravitino. 
For a theoretically consistent and natural realization 
 of such a mass spectrum as well as neutralino dark matter 
 physics in the setup, see Ref.~\cite{emergentSUSY}}.
Because of the theoretical consistency mentioned above, 
 we constrain the deflection parameter not to exceed $\mathcal{O}(1)$ 
 and reconsider the phenomenology of the positively deflected AMSB
 scenario. 
In this paper we perform the parameter scan 
 by taking into account a variety of phenomenological constraints. 
In particular, the relic abundance of the neutralino dark matter 
 and the observed Higgs boson mass play the crucial roles 
 in identifying the allowed parameter region.

%- Structure of the paper

The paper is organized as follows.
We give a brief review of the deflected AMSB scenario in Sec. 2. 
In Sec. 3, we perform the parameter scan of 
 the positively deflected AMSB scenario 
 by taking into account various phenomenological constraints 
 and identify the allowed parameter region. 
We also show the benchmark mass spectra 
 for the parameter sets from the allowed region. 
For these benchmark points, we calculate the elastic scattering 
 cross section of the dark matter neutralino with nuclei 
 and discuss its implication for future dark matter 
 detection experiments. 
The last section is devoted to conclusions.

%%%%%%%%%%%%%%%%%%%%%%%%%%%%%%%%%%%%%%
\section{Deflected anomaly mediation}
%%%%%%%%%%%%%%%%%%%%%%%%%%%%%%%%%%%%%%

%- AMSB revisited
%- Deflected AMSB revisited

Assuming the sequestering between the hidden 
 and visible sectors \cite{AMSB1}, the direct SUSY-breaking 
 mediation from the hidden sector to the visible one 
 is forbidden. 
However, it was found that in the context of supergravity, 
 there always exists the SUSY-breaking mediation 
 through the superconformal anomaly, namely, 
 the AMSB \cite{AMSB1, AMSB2}.
In this scenario, sfermion squared masses are predicted 
 to be proportional to the beta-function coefficients 
 of the MSSM gauge coupling RG equations, so that 
 slepton squared masses are negative. 
In order to solve this tachyonic slepton problem, 
 we introduce the messenger sector similar to that in the minimal GMSB \cite{GMSB}. 
The corresponding superpotential is given by 
\begin{eqnarray}
 W = \sum_{i=1}^N S \bar{\Psi}_i \Psi^i , \label{messenger}
\end{eqnarray}
where $S$ is a singlet chiral superfield, 
 and $N$ is the number of vectorlike pairs of messengers 
 $\Psi^i$ and $\bar{\Psi}_i$ in the fundamental and antifundamental 
 representations under the MSSM gauge groups. 
Here we have used the $SU(5)$ gauge group notation 
 for the MSSM gauge group, for simplicity, 
 under which the messengers are the vectorlike pairs 
 of ${\bf 5}+{\bf 5^*}$ representation.

Once the nonzero VEVs of the scalar and $F$-components of $S$ 
 are developed, the messenger sector gives rise to the GMSB-like 
 contributions to the soft SUSY-breaking terms, which can be 
 represented as 
\begin{eqnarray}
  \frac{ F_S }{ S } = 
  d F_\phi, 
\label{deflection}  
\end{eqnarray} 
where the coefficient $d$ is the deflection parameter. 
In the deflected anomaly mediation, 
 $ F_S $ is generated 
 by the primary SUSY-breaking source $F_\phi$. 
In a simple scenario, the deflection parameter is evaluated 
as \cite{positiveAMSB} 
\begin{eqnarray} 
 d  \sim  - 2  
\frac{ \frac{\partial W}{\partial S} }
{S \frac{\partial^2 W}{\partial S^2}}
\end{eqnarray} 
from the superpotential of the singlet superfield $W(S)$.  
Since this SUSY breaking in the messenger sector 
 is a secondary SUSY-breaking, the theoretical consistency 
 requires $|d| \lesssim {\cal O}(1)$, in other words, 
 the secondary SUSY breaking order parameter cannot be 
 much greater than the primary SUSY-breaking scale $F_\phi$.

By using the method established in Ref. \cite{SUSYmethod}, 
 the soft SUSY-breaking terms can be extracted from 
 the renormalized gauge couplings and the supersymmetric wave-function 
 renormalization coefficients as follows \cite{negativeAMSB, positiveAMSB}:
\begin{eqnarray}
\frac{M_i(\mu)}{\alpha_i(\mu)} 
 &=& 
\frac{F_\phi}{2} \left( \frac{\partial}{\partial \ln \mu} 
- d \frac{\partial}{\partial \ln |S|} \right) \frac{1}{\alpha_i(\mu,S)} , \label{soft1} \\
m_I^2(\mu) & = &
 - \frac{|F_\phi|^2}{4} \left
( \frac{\partial}{\partial \ln \mu} - d \frac{\partial}{\partial \ln
|S|} \right)^2 \ln Z_I(\mu,S) , 
\label{soft3} \\
A_I(\mu) & = &
- \frac{F_\phi}{2} \left( \frac{\partial}{\partial \ln \mu} 
- d \frac{\partial}{\partial \ln |S|} \right) \ln Z_I(\mu,S) , 
\label{soft2} 
\end{eqnarray}
where the gauge couplings and the wave-function renormalization
 coefficients are given by
\begin{eqnarray}
\alpha_i^{-1} (\mu,S) & = &
\alpha_i^{-1}(\Lambda_{cut}) + \frac{b_i - N}{4\pi} 
\ln \frac{S^\dagger S}{\Lambda_{cut}^2} + \frac{b_i}{4\pi} \ln
\frac{\mu^2}{S^\dagger S} , 
\label{gaugecouplings} \\
Z_I (\mu, S) & = &
\sum_{i=1,2,3} Z_I(\Lambda_{cut}) 
\left( \frac{\alpha(\Lambda_{cut})}{\alpha(S)} \right)^\frac{2c_i}{b_i - N}
 \left( \frac{\alpha(S)}{\alpha(\mu)} \right)^\frac{2c_i}{b_i} . 
\label{wavefunction}
\end{eqnarray}
The index $i=1,2,3$ corresponds to the MSSM gauge group 
 $U(1)_Y \times SU(2)_L \times SU(3)_C$, 
 and $b_i=\{-33/5, -1, 3 \}$ ($i=1,2,3$) 
 are the beta-function coefficients of the MSSM gauge coupling RG equations. 
The messenger scale $M_\text{Mess} = S $ 
 plays a role of the intermediate threshold 
 between the UV cutoff $\Lambda_{cut}$ and 
 the electroweak scale.

%- Boundary conditions in positively deflected AMSB:

Substituting Eqs.~(\ref{gaugecouplings}) and (\ref{wavefunction}) 
 into Eqs.~(\ref{soft1}) and (\ref{soft3}), 
 we obtain the solutions for the RG equations of the soft terms. 
At the messenger scale, 
 the MSSM gaugino masses are given by 
\begin{eqnarray}
M_i (M_\text{Mess}) = - \frac{\alpha_i}{4\pi} 
  F_\phi \left( b_i + d N \right).   
\label{soft-gaugino}
\end{eqnarray}
For the $A$ parameters of the third-generation, 
 we have 
\begin{eqnarray}
A_t (M_\text{Mess})    & = & 
- \frac{F_\phi}{(4\pi)^2} 
\left( 6|Y_t|^2 + |Y_b|^2 - \frac{16}{3} g_3^2 - 3g_2^2 - \frac{13}{15} g_1^2 \right) , \\
	A_b (M_\text{Mess})    & = & 
			- \frac{F_\phi}{(4\pi)^2} \left( |Y_t|^2 + 6|Y_b|^2 + |Y_\tau|^2 - \frac{16}{3} g_3^2 - 3 g_2^2 - \frac{18}{5} g_1^2 \right) , \\
	A_\tau (M_\text{Mess}) & = & 
			- \frac{F_\phi}{(4\pi)^2} \left( 3|Y_b|^2 +
4|Y_\tau|^2 - 3 g_2^2 - \frac{9}{5} g_1^2 \right) , 
\end{eqnarray}
 where $Y_{t,b,\tau}$ are the Yukawa couplings 
 of the third generation quarks and lepton. 
Finally, the sfermion squared masses are given by 
\begin{eqnarray}
m_{H_u}^2 (M_\text{Mess})
& = & 	m_{H_d}^2 (M_\text{Mess}) = 
F_\phi^2 \left[
\frac{3}{10} \left( \frac{\alpha_1}{4\pi} \right)^2 G_1 
+ \frac{3}{2} \left( \frac{\alpha_2}{4\pi} \right)^2 G_2 \right] , \\
m_{\tilde{L}}^2 (M_\text{Mess}) & = & 
F_\phi^2 \left[\frac{3}{10} \left( \frac{\alpha_1}{4\pi} \right)^2 G_1 
+ \frac{3}{2} \left( \frac{\alpha_2}{4\pi} \right)^2 G_2 \right] , \\
m_{\tilde{E}}^2 (M_\text{Mess}) & = & 
F_\phi^2 \left[\frac{6}{5} \left( \frac{\alpha_1}{4\pi} \right)^2  G_1
 \right], \\
m_{\tilde{Q}}^2 (M_\text{Mess}) & = & 
F_\phi^2 \left[
\frac{1}{30} \left( \frac{\alpha_1}{4\pi} \right)^2 G_1 
+ \frac{3}{2} \left( \frac{\alpha_2}{4\pi} \right)^2 G_2 
+ \frac{8}{3} \left( \frac{\alpha_3}{4\pi} \right)^2 G_3 \right] , \\
m_{\tilde{U}}^2 (M_\text{Mess}) & = &
F_\phi^2 \left[
\frac{8}{15} \left( \frac{\alpha_1}{4\pi} \right)^2 G_1 
+ \frac{8}{3} \left( \frac{\alpha_3}{4\pi} \right)^2 G_3 \right] , \\
m_{\tilde{D}}^2 (M_\text{Mess}) & = &
F_\phi^2 \left[
 \frac{2}{15} \left( \frac{\alpha_1}{4\pi} \right)^2 G_1 
 + \frac{8}{3} \left( \frac{\alpha_3}{4\pi} \right)^2 G_3\right] ,
\end{eqnarray}
where 
\begin{eqnarray}
	G_i = N d^2 + 2 N d + b_i.  
\end{eqnarray}
These soft terms are used as the boundary conditions 
 for the RG-equation evolution of MSSM soft terms in our analysis. 
Note that the limit $d=0$ reproduces the pure AMSB result
 for the soft terms, 
 while the limits $d \to \infty$ and $F_\phi \to 0$ 
 (while keeping $d F_\phi$ finite)
 leads to the soft terms in the GMSB scenario.

%%%%%%%%%%%%%%%%%%%%%%%%%%%%%%%%%%%%%%%%%%%%%%%%%%%%%%%
\section{
Numerical analysis
}
%%%%%%%%%%%%%%%%%%%%%%%%%%%%%%%%%%%%%%%%%%%%%%%%%%%%%%%%

%- Process of calculation
%- Free parameter set

In the deflected AMSB scenario, the soft terms are characterized 
 by 5 free parameters, 
\begin{eqnarray}
 N, \; d, \; M_\text{Mess}, \; F_\phi, \; \tan \beta ,
\end{eqnarray} 
and one sign of the $\mu$ parameter. 
In this paper, we only consider $\text{sign} (\mu) = +$, 
 which gives rise to a positive contribution of sparticles 
 to the muon anomalous magnetic moment. 
For various fixed values of $N$, $d$ and $\tan \beta$, 
 we scan over the other free parameters $M_\text{Mess}$ and $F_\phi$ 
 and identify a parameter region which is consistent 
 with a variety of phenomenological constraints. 
For numerical analysis, we employ the SOFTSUSY package 
 version 3.3.1 \cite{softsusy} to solve the MSSM RG equations  
 and compute the mass spectrum, with the inputs of the gaugino 
 masses, $A$ parameters and the sfermion squared masses 
 at the messenger scale given in the previous section. 
The micrOMEGAs package version 2.4.5 \cite{micromega} 
 is used to calculate the neutralino dark matter relic 
 abundance and other phenomenological constraints 
 with the output of SOFTSUSY.

%- Figures of parameter scanning

\begin{figure}[ht]
\begin{center}
\scalebox{1.5}[1]{
\includegraphics[scale=1.0]{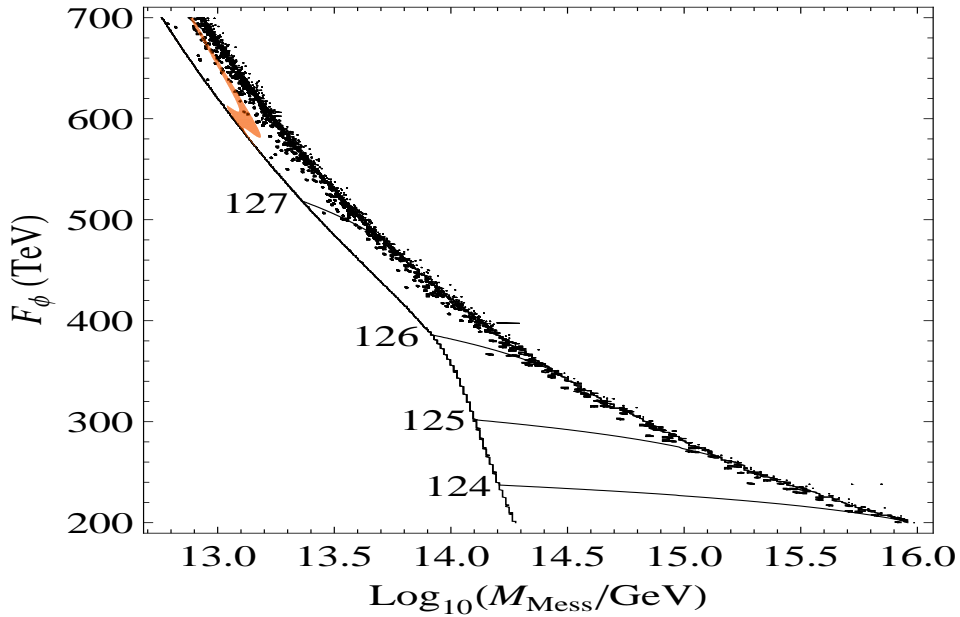}
}
\caption{
Results for $N=1$, $d=2$, and $\tan \beta = 10$. 
Various values of the resultant SM-like Higgs boson mass 
 are shown as the contours. 
There are two diagonal boundaries, outside of which 
 values are theoretically excluded. 
The red strip indicates the region where 
 the relic abundance of neutralino dark matter 
 is consistent with observation. 
} 
\label{N1d2tan10}
\end{center}
\end{figure}

\begin{figure}[ht]
\begin{center}
\scalebox{1.5}[1]{
\includegraphics[scale=1.0]{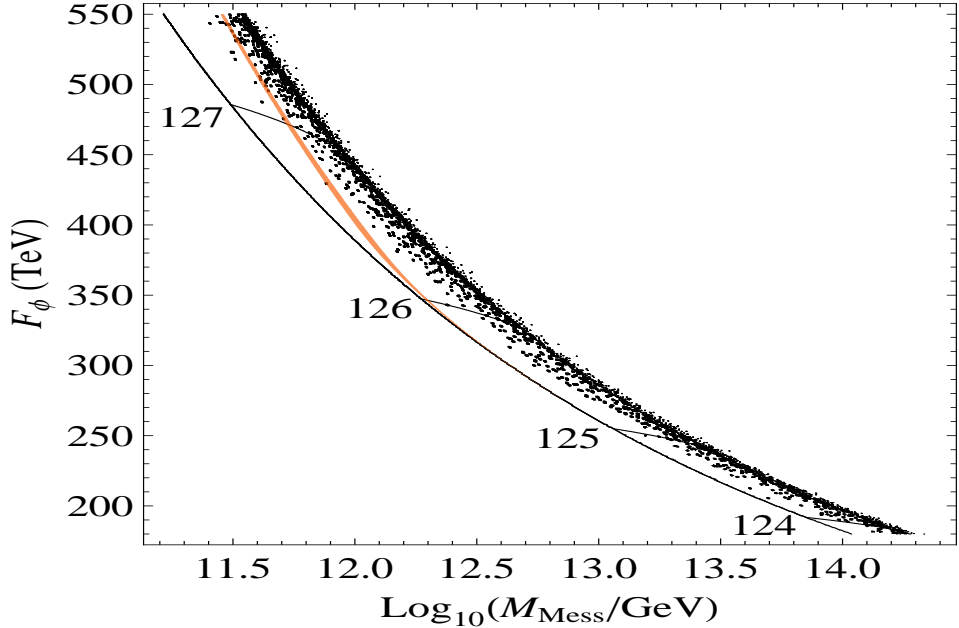}
}
\caption{
The same as in Fig.~\ref{N1d2tan10}, 
 but for $N=2$, $d=1.5$, and $\tan \beta = 10$. 
} 
\label{N2d1.5tan10}
\end{center}
\end{figure}

\begin{figure}[ht]
\begin{center}
\scalebox{1.5}[1]{
\includegraphics[scale=1.0]{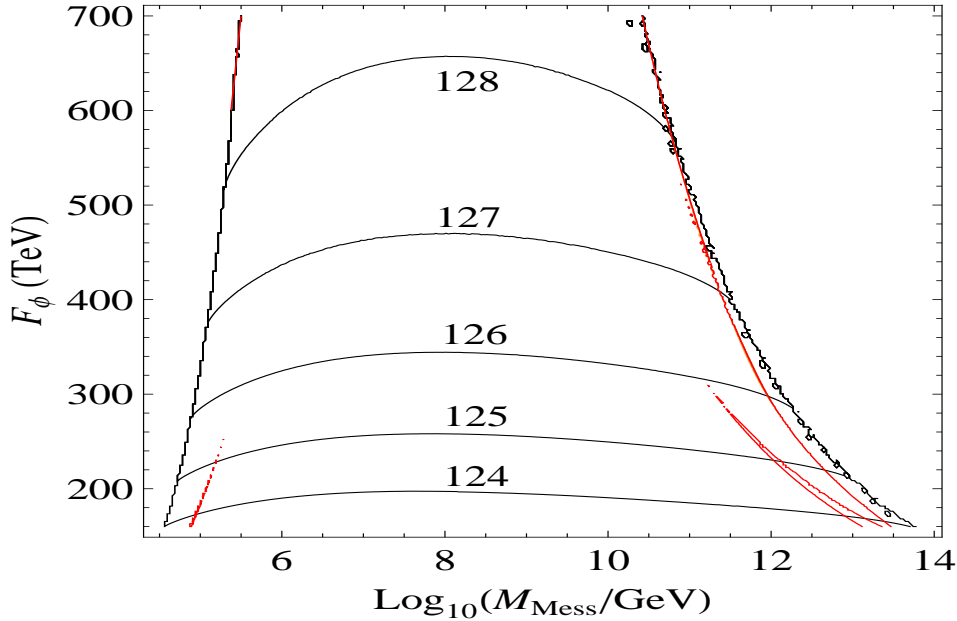}
}
\caption{
Results for $N=2$, $d=2$, and $\tan \beta = 10$. 
Various values of the resultant SM-like Higgs boson mass 
 are shown as the contours. 
There are two diagonal boundaries, outside of which values 
 are theoretically excluded. 
The four red strips indicate the region 
 where the relic abundance of neutralino dark matter 
 is consistent with observation. 
} 
\label{N2d2tan10}
\end{center}
\end{figure}

\begin{figure}[ht]
\begin{center}
\scalebox{1.5}[1]{
\includegraphics[scale=1.0]{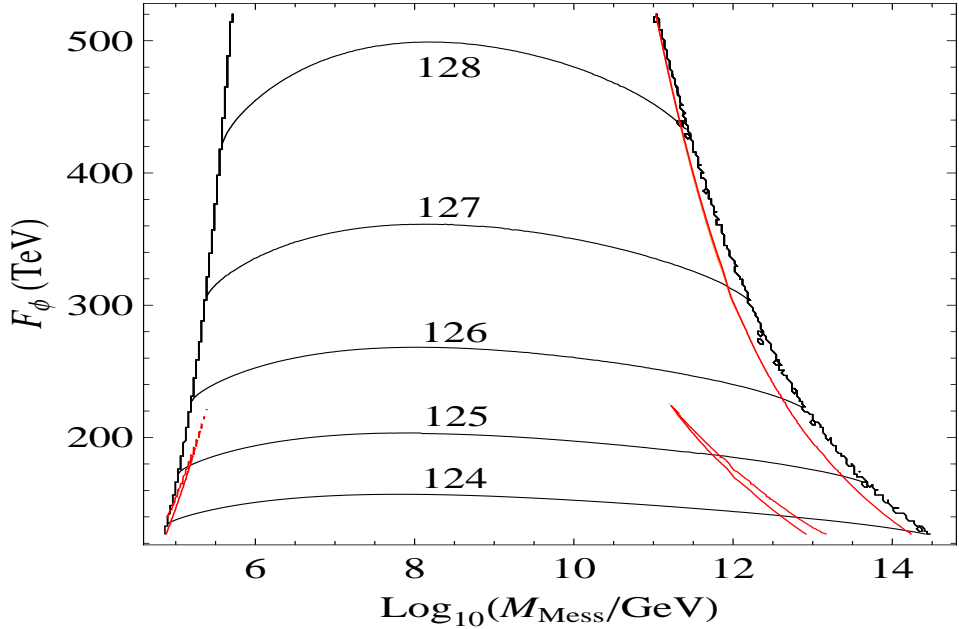}
}
\caption{
The same as in Fig.~\ref{N2d2tan10} 
 but for $\tan \beta = 20$. 
} 
\label{N2d2tan20}
\end{center}
\end{figure}

\begin{figure}[ht]
\begin{center}
\scalebox{1.5}[1]{
\includegraphics[scale=1.0]{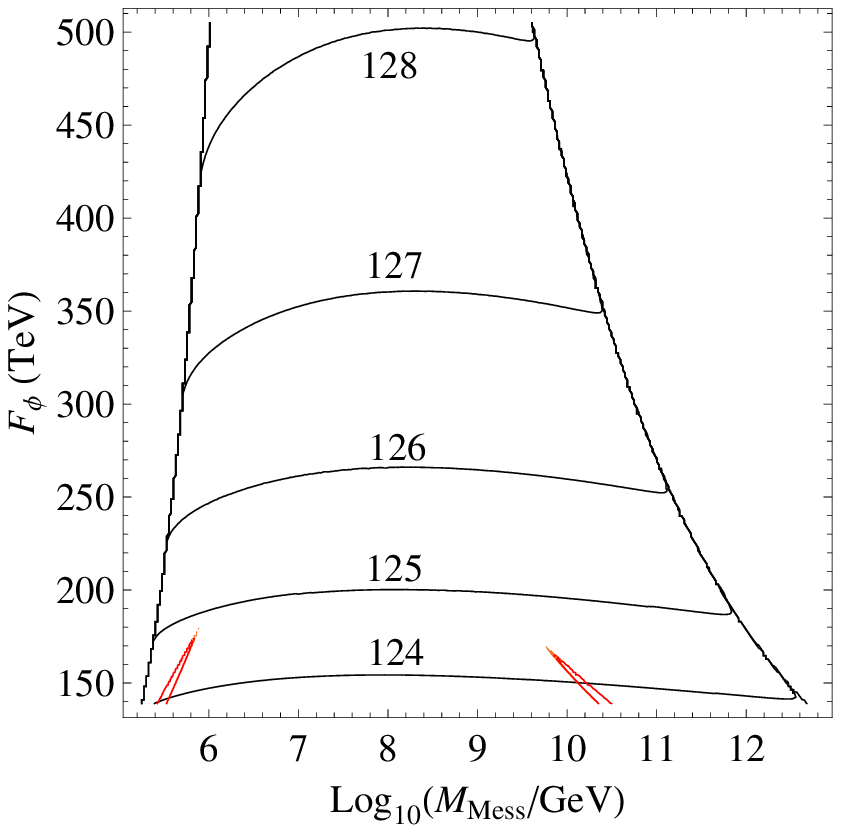}
}
\caption{
The same as in Fig.~\ref{N2d2tan10} 
 but for $\tan \beta = 30$. 
} 
\label{N2d2tan30}
\end{center}
\end{figure}

The results are shown in Figs. \ref{N1d2tan10}-\ref{N2d2tan30}.  
%, \ref{N2d1.5tan10}, \ref{N2d2tan10}, \ref{N2d2tan20}
%
The various values of the resultant SM-like Higgs boson mass 
 are depicted as contours. 
Since the soft terms at the messenger scale are proportional to 
 $F_\phi$, the sparticle masses (except Higgsino masses) 
 scale as $F_\phi$. 
As a result, the SM-like Higgs boson mass, 
 which is mainly controlled by stop masses and $A_t$, 
 becomes heavier as $F_\phi$ is raised. 
For the parameters in the red regions, 
 the relic density of the neutralino dark matter 
 can be consistent with the observed data \cite{WMAP}:
\begin{eqnarray}
\Omega_\text{CDM} h^2 = 0.1120 \pm 0.0056 . 
\label{CDM}
\end{eqnarray}

%- Properties of LSP in each figure
%- Reasons for boundaries of each figure (black dots and zigzag lines)

%%%%%%%%%%%%%%%%%%%%%%%%%%%%%%%%%%%%%%%%%%%%%%%
%%%%%%%%%%%%%%%%%%%%%%%%%%%%%%%%%%%%%%%%%%%%%%%

Figure ~\ref{N1d2tan10} shows the results 
 for $N=1$, $d=2$ and $\tan \beta = 10$. 
The region inside the left and right diagonal boundaries 
 is theoretically allowed. 
In the region outside of the left boundary 
 the LSP is found to be the lighter stau 
 and is excluded due to the electrically charged LSP.
Since there appear many dots, the right boundary is not clear, 
 compared to the left one. 
This is because around that region, the convergence of 
 our numerical analysis with the SOFTSUSY turns out 
 to be very sensitive to the input parameters 
 and we have omitted the parameter sets which 
 do not converge in the SOFTSUSY analysis. 
In the region outside of the right boundary, 
 the radiative electroweak symmetry breaking 
 is not achieved (no-EWSB region) and 
 hence the region is excluded.

For $N=1$ and $d=2$, we always find $M_2 < M_1$ 
 (see Eq.~(\ref{soft-gaugino})), and the LSP neutralino 
 will be either wino-like, Higgsino-like, or the mixture of both. 
In the red region around $F_\phi=600$ TeV, we have found 
 the wino-like LSP neutralino, while the Higgsino-like neutralino 
 is found along the thin strip for $F_\phi > 600$ TeV. 
For this region, the Higgs boson mass is predicted as $m_h > 127$ GeV, 
 so that the region reproducing the observed dark matter relic density 
 cannot be compatible with the Higgs boson mass measured by ATLAS and CMS. 
We have found that this conclusion remains the same for 
 the $N=1$ case with various values of $d$ and $\tan \beta$.

%%%%%%%%%%%%%%%%%%%%%%%%%%%%%%%%%%%%%%%%%%%%%%%
%%%%%%%%%%%%%%%%%%%%%%%%%%%%%%%%%%%%%%%%%%%%%%%

In Figure \ref{N2d1.5tan10}, we have examined the case 
 with $N = 2$, $d = 1.5$ and $\tan \beta = 10$.
Similarly to Fig.~\ref{N1d2tan10}, 
 there are two boundaries and the regions outside 
 of them are excluded by the same conditions. 
Again, the figure shows many dots by the same reason 
 as in Fig.~\ref{N1d2tan10}. 
The difference is that in this case the LSP neutralino 
 cannot be wino-like as can be understood from Eq.~(\ref{soft-gaugino}). 
We have found that along the thin red strip, 
 the observed dark matter abundance is reproduced 
 with the Higgsino-like LSP neutralino. 
As $F_\phi$ becomes larger along the red strip, 
 the LSP neutralino become slightly heavier 
 and the lighter chargino mass becomes closer 
 to the LSP neutralino mass. 
Note that the parameters on the red strip with 
 $270\; \text{TeV} \lesssim F_\phi \lesssim 450 \; \text{TeV}$ 
 predict a Higgs boson mass consistent with the LHC data. 
When the red region is very close to the left boundary, 
 the lighter stau is very much degenerate with 
 the LSP neutralino. 
However, the coannihilation process of the LSP neutralino 
 with the stau is not important 
 because its relic abundance is determined dominantly by 
 the neutralino pair annihilation process to 
 $W^\pm$ with the charginos exchanged in the $t$-channel. 
Although the sparticle mass spectrum is quite different, 
 phenomenology of the parameters in the red region 
 is similar to the so-called focus point region  
 in the constrained MSSM, 
 where Higgsino-like neutralinos and charginos 
 are light and the others are quite heavy.

%%%%%%%%%%%%%%%%%%%%%%%%%%%%%%%%%%%%%%%%%%%%%%%
%%%%%%%%%%%%%%%%%%%%%%%%%%%%%%%%%%%%%%%%%%%%%%%

Figure \ref{N2d2tan10} shows the results 
 for $N=2$, $d=2$ and $\tan \beta=10$. 
Comparing this figure with the previous one, 
 we see a significant movement of the left boundary 
 to the left due to the increase of the deflection 
 parameter $d=1.5 \to 2$. 
The conditions that define the left boundary are a little involved. 
For values of $F_\phi \gtrsim 600$ TeV 
 the left boundary is specified by the no-EWSB condition, 
 while for the parameters outside of the boundary 
 (with $F_\phi \lesssim 600$ TeV) 
 the CP-odd heavy Higgs boson is found to be tachyonic. 
The right boundary is specified by the no-EWSB condition  
 as in the previous figures.

The thin red strips show the allowed regions 
 for the relic abundance of the LSP neutralino
 to be consistent with observation. 
For parameters on the strip close to the left boundary 
 for $F_\phi \gtrsim 600$ TeV and on the strip 
 along the right boundary, the LSP neutralino 
 is found to be Higgsino-like.  
As before, this region is similar to the focus point region 
 in the constrained MSSM. 
On the other red strips, we have found the bino-like neutralino 
 whose mass is close to a half of the heavy Higgs boson masses. 
Thus, the correct dark matter abundance is achieved 
 by the $s$-channel resonance via the heavy Higgs bosons 
 in the neutralino pair annihilation process. 
These regions correspond to the so-called funnel region 
 in the constrained MSSM. 
Now we see that three separate regions satisfy 
 the conditions of the dark matter relic abundance 
 and the Higgs boson mass simultaneously, 
 namely, two funnel-like regions and one focus point-like region.

%%%%%%%%%%%%%%%%%%%%%%%%%%%%%%%%%%%%%%%%%%%%%%%
%%%%%%%%%%%%%%%%%%%%%%%%%%%%%%%%%%%%%%%%%%%%%%%

%%%%%%%%%%%%%%%%%%%%%%%%%%%%%%%%%%%%%%%%%%%%%%%
%%%%%%%%%%%%%%%%%%%%%%%%%%%%%%%%%%%%%%%%%%%%%%%

In the following, we keep the values of $N=2$ and $d=2$, 
 but raise $\tan \beta$ to $20$ and $30$. 
The results are shown in Figs.~\ref{N2d2tan20} and \ref{N2d2tan30}, 
 respectively. 
Figure \ref{N2d2tan20} looks identical to Figure \ref{N2d2tan10} 
 and in fact, the physics for the allowed regions are the same. 
However, the left boundary shifts to the right 
 and hence the region inside the two boundaries 
 becomes narrower by the increase of $\tan \beta$. 
We also see that the funnel-like strips move downward. 
It can be seen in Figure \ref{N2d2tan30} that the boundaries 
 shift further (the left boundary shifts to the right and the right boundary shifts to the left) and make the region between them narrower, 
 and the funnel-like strips move further downward. 
The left boundary in Figure \ref{N2d2tan30} is specified 
 by the condition of the tachyonic CP-odd Higgs boson, 
 while the right boundary is given by the tachyonic 
 charged Higgs boson, and not by the no-EWSB condition. 
The red strips along the right boundary 
 in Figs.~\ref{N2d2tan10} and \ref{N2d2tan20}
 disappears in Fig.~\ref{N2d2tan30}. 
In this figure, the regions consistent with 
 both the dark matter relic abundance and the Higgs boson mass
 appear only on the funnel-like strips. 
When we increase $\tan \beta$ further, we find that the predicted Higgs
 boson mass in the funnel-like region becomes too small 
 to reproduce the observed Higgs boson mass.

%- Benchmark points and other phenomenological constraints.

%%%%%%%%%%%%%%%%%%%%%%%%%%%%%%%%%%%%%%%%%%%%%%%

\begin{table}
\caption{Benchmark particle mass spectra 
 for the case $N=2$ and $\tan \beta = 10$. 
Masses of particles are given in GeV.
The values of the branching fractions of $b \rightarrow s + \gamma$, $B_{s} \rightarrow \mu^{+} \mu^{-}$, and $B \rightarrow \tau \nu_\tau$, the muon anomalous magnetic dipole moment $\Delta a_\mu$, and the neutralino relic density are calculated for each benchmark point.
The spin-independent and spin-dependent cross sections for the neutralino-proton elastic scattering are also provided.} 
\label{benchmark}
\begin{center}
\scalebox{1}[1]{
\begin{math} 
\begin{array}{|c||c|ccc|}
\hline

d & 1.5 &  2 &  2 &  2		\\
F_\phi 	&  2.800 \times 10^{5} 	&   2.500 \times 10^{5} 	
   &  2.500 \times 10^{5} & 2.500 \times 10^{5} 	\\
M_\text{Mess} 	&  6.457 \times 10^{12} &  1.902 \times 10^{5}
  &  6.220 \times 10^{11} &   2.176 \times 10^{12} 	\\
\hline 

h^0 	&  125.3 	&   125.3 	&  125.3 	&      125.4 	\\

H^0 	&  1389 	&   1850 	&  1921 	&      1595 	\\

A^0 	&  1389 	&   1849 	&  1921 	&      1595 	\\

H^\pm 	&  1391	 	&   1851 	&  1923 	&      1597 	\\

\tilde{g} &  10106 	&   10651 	&  10519 	&      10512 	\\
  
\tilde{\chi}^0_{1,2} 	&  852.7,  858.5	&  918.7,  1818	&  943.2,  1480	& 900.1,  912.9 \\

\tilde{\chi}^0_{3,4} 	&  1422,  1442 		&  1841,  2003 	&  1490,  1918 	& 962.8,  1915 \\

\tilde{\chi}^\pm_{1,2} 	&  853.9, 1421 	&  1819, 2003 
  &  1480, 1918  & 913.6, 1914 \\

\tilde{u}, \tilde{c}_{R,L}	&  9839, 10200 	&  10522, 10947 &  10303, 10750 & 10302, 10748 \\
   
\tilde{d}, \tilde{s}_{R,L}	&   9703, 10200 &  10494, 10947 &  10170, 10750 & 10152, 10748 \\

\tilde{t}_{1,2}				&  7865, 9297 	&  9951,  10670 &  8377, 9872  	& 8274, 9827 \\
    
\tilde{b}_{1,2} 			&  9296,  9669 	&  10478, 10670	&  9870, 10137  & 9826, 10118  \\    

\tilde{\nu}^{e, \mu}_L &  3408 &  3190 &   3761 &  3829 	 \\

\tilde{e}, \tilde{\mu}_L &  3410,  3408 &  3191, 3190 
&   3762,  3760 & 3830, 3828 \\
     
\tilde{e}, \tilde{\mu}_R &  914.3, 914.1 &  1284,  1284  
&  1404, 1404  & 1420, 1420 \\
   
\tilde{\nu}^{\tau}_L 	&  3400 &   3188 & 3754 &     3821 	 \\
 
\tilde{\tau}_{1,2} &  853.0,  3401 &  1277,  3189  
&  1368,  3755  & 1382, 3822 \\

\hline

%\Delta \rho 	& 5.81 \times 10^{-07}	& 5.49 \times 10^{-07} 	
%&  5.24 \times 10^{-07} &   5.18 \times 10^{-07} 	\\

\text{BR}(b \rightarrow s + \gamma) 
				&  3.44 \times 10^{-4} & 3.38 \times 10^{-4} 	&  3.37 \times 10^{-4} &       3.41 \times 10^{-4} 	\\

\text{BR}(B_{s} \rightarrow \mu^{+} \mu^{-} ) 
				&  3.06 \times 10^{-9} & 3.06 \times 10^{-9} 	&  3.06 \times 10^{-9} &     3.06 \times 10^{-9} 	\\

\frac{\text{BR}^\text{exp}(B \rightarrow \tau \nu_\tau)}{\text{BR}^\text{SM}(B \rightarrow \tau \nu_\tau)}
&  0.997 	& 0.998 	&  0.999 	& 0.998 	\\

\Delta a_\mu 	&  3.98 \times 10^{-11} & 2.42 \times 10^{-11} 	&
2.38 \times 10^{-11} &         2.62 \times 10^{-11} 	\\
\hline

\Omega h^2 &  0.1121 	& 0.1121 &  0.1121 & 0.1121 	\\
\hline
\sigma_\text{SI}^{\chi-\text{p}} (\text{pb})
& 4.498 \times 10^{-9}	& 1.426 \times 10^{-11}	& 8.047 \times
10^{-11} & 1.257 \times 10^{-8}		\\

\sigma_\text{SD}^{\chi-\text{p}} (\text{pb})	
& 1.325 \times 10^{-6}	& 2.189 \times 10^{-8}	&8.283 \times 10^{-8}	
& 1.025 \times 10^{-5}	\\
\hline
\end{array}
\end{math}}
\end{center}
\end{table}

%%%%%%%%%%%%%%%%%%%%%%%%%%%%%%%%%%%%%%%%%%%%%%%

To see the particle mass spectrum, 
 we list four benchmark points in Table~\ref{benchmark}, 
 one from Fig.~\ref{N2d1.5tan10} and 
 the other three from Fig.~\ref{N2d2tan10}, 
 which simultaneously satisfy the constraints 
 on the neutralino dark matter relic abundance 
 and the measured SM-like Higgs boson mass. 
In addition to the constraints, 
 we also take into account other phenomenological constraints:  
 the branching ratios of 
	$b \rightarrow s \gamma$ \cite{bsgamma}, 
%	$B_{s} \rightarrow \mu^{+} \mu^{-} $ \cite{bsmumu},
	$B \rightarrow \tau \nu_\tau$ \cite{btaunu},
 and the muon anomalous magnetic moment 
	$a_{\mu} = (g_{\mu} - 2)/2$ \cite{gmu},
\begin{eqnarray}
&	2.85 \times 10^{-4} \leq \text{BR}(b \rightarrow s + \gamma) \leq 4.24
\times 10^{-4} \; (2 \sigma ) , &  \label{bsg} \\
%& 2.0 \times 10^{-9} <	\text{BR}(B_{s} \rightarrow \mu^{+} \mu^{-} ) < 4.7 \times 10^{-9} \; (3.5 \sigma),  & \label{bsmm} \\
&	\dfrac{\text{BR}^\text{exp}(B \rightarrow \tau \nu_\tau)}{\text{BR}^\text{SM}(B \rightarrow \tau \nu_\tau)} = 1.25 \pm 0.40, & \label{btn} \\
&	\Delta a_{\mu} = a_\mu^\text{exp} - a_\mu^\text{SM} = (26.1 \pm 8.0) \times 10^{-10} \;
(3.3 \sigma ) .  & 
\label{delta-a} 
\end{eqnarray}
In particular, we consider the most recent limits of the decay process $B_{s} \rightarrow \mu^{+} \mu^{-} $ announced by the LHCb Collaboration \cite{bsmumu}:
\begin{eqnarray}
& 2.0 \times 10^{-9} <	\text{BR}(B_{s} \rightarrow \mu^{+} \mu^{-} ) < 4.7 \times 10^{-9} \; (3.5 \sigma),  & \label{bsmm}
\end{eqnarray}
We can see that all the above phenomenological constraints, 
 except for $\Delta a_{\mu}$, are well satisfied. 
The SUSY contribution to the muon anomalous magnetic dipole moment 
 is too small to explain the discrepancy between 
 the observed value and the SM prediction. 
This is because in order to reproduce the SM-like Higgs boson mass 
 around 125 GeV, the particles are all found to be heavy. 
This actually happens in most of the well-known SUSY models 
 \cite{ImplicationHiggs}.

In Table~\ref{benchmark}, 
 we see that only the neutralino(s) and chargino can be 
 lighter than 1 TeV. 
The first column corresponds to a point in Fig.~\ref{N2d1.5tan10}, 
 where the LSP neutralino is Higgsino-like. 
The second and third columns correspond to 
 2 points in the funnel-like regions in Fig.~\ref{N2d2tan10}, 
 where the LSP neutralinos are bino-like. 
The last column corresponds to a point in Fig.~\ref{N2d1.5tan10}, 
 where the LSP neutralino is Higgsino-like, as in the first column.

In Table \ref{benchmark}, we also list the prediction of the spin-independent (SI) 
 and the spin-dependent (SD) cross sections for neutralino 
 elastic scattering off a proton, which are relevant to 
 the direct/indirect dark matter detection experiments. 
When the LSP neutralino is bino-like, 
 both the cross sections are very small and 
 beyond the sensitivity of future experiments. 
On the other hand, for the Higgsino-like neutralino 
 in the first and last columns, 
 we have found $\sigma_{SI} = {\cal O}(10^{-8})$ 
 and $\sigma_{SD} = {\cal O}(10^{-6}-10^{-5})$. 
Future direct dark matter search experiments, 
 such as XENON1T \cite{ZENON1T}, can cover the SI cross 
 section up to $ \sigma_{SI} = {\cal O}(10^{-10})$  
 for a dark matter particle with mass 1 TeV, 
 so that the neutralino dark matter in our scenario 
 can be tested.

%%%%%%%%%%%%%%%%%%%%%%%%%%%%%%
\section{Conclusions}
%%%%%%%%%%%%%%%%%%%%%%%%%%%%%%
In the positively deflected AMSB, 
 the tachyonic slepton problem in the pure AMSB is ameliorated 
 by introducing the messenger sector that brings 
 the GMSB-like contributions to lift up the slepton squared masses 
 to be positive. 
In this scenario, the lightest neutralino can be the LSP 
 (as usual) and hence the dark matter candidate. 
In the light of the recent discovery of the Higgs boson at the LHC, 
 we have reconsidered this scenario, 
 in particular the phenomenology of the neutralino dark matter  
 with natural values of the deflection parameter. 
We have also taken into account other phenomenological constraints. 
By fixing $N$, $d$ and $\tan \beta$, 
 we have performed the parameter scan with various values 
 of $M_\text{Mess}$ and $F_\phi$ and identified the parameter regions 
 that simultaneously satisfy the constraints 
 on the dark matter relic abundance and the observed (SM-like)
 Higgs boson mass. 
We have found that in most of the allowed parameter regions, 
 the dark matter neutralino is Higgsino-like. 
The four benchmark points for the particle spectrum 
 are listed in Table \ref{benchmark}, for which all phenomenological constraints,  
 except the muon anomalous magnetic moment, 
 are well satisfied. 
We have found that although the particle mass spectrum is very high 
 and the SUSY search at the LHC is quite challenging, 
 the SI cross section of the Higgsino-like dark matter 
 stays within the reach 
 of future dark matter direct detection experiments. 
Thus, such experiments can reveal the signature of SUSY.

Finally, we comment on the following two issues. 
First, since our resultant sparticle mass spectrum is very high, 
 our scenario needs a fine-tuning for the $\mu$ parameter 
 to obtain the correct electroweak scale. 
For the benchmark points in Table \ref{benchmark}, 
 we find $\mu \simeq 900$ GeV, so that 
 the level of fine-tuning is about 0.5 \%, 
 when we estimate it by $\dfrac{m_Z^2}{2 \mu^2}$, 
 with $m_Z$ being the $Z$ boson mass.
Second, the SM-like Higgs boson can potentially yield 
 deviations in its properties from those of the SM Higgs boson. 
Although the Higgs boson properties measured at the LHC 
 are mostly consistent with the SM predictions, 
 the signal strength of the diphoton decay mode 
 shows about a 2$\sigma$ discrepancy between the observed value 
 and the SM expectation \cite{atlas, cms} 
 (see also Refs. \cite{Higgs-diphoton1, Higgs-diphoton2}). 
This deviation could be an indirect signal of sparticles. 
For the benchmark points, we have calculated 
 the signal strength for the channel $gg \to h^0 \to \gamma \gamma$ 
 and compared it with the SM prediction. 
Using the FeynHiggs package \cite{FeynHiggs}
 with the output of SOFYSUSY, we have found that 
 the deviation is about a few percent and hence negligible. 
This is also because of the heavy sparticle mass spectrum.

%%%%%%%%%%%%%%%%%%%%%%%%%%%%%%
\section*{Acknowledgment}
%%%%%%%%%%%%%%%%%%%%%%%%%%%%%%
The work of N.O. is supported in part by the DOE Grants, 
 No. DE-FG02-10ER41714.
H.M.T. would like to thank The Theoretical Group, 
 University of Lyon 1 (especially Aldo Deandrea) 
 for hospitality and the LIA-FVPPL project 
 for financial support during his visit.

%%%%%%%%%%%%%%%%%%%%%%%%%%%%%%%%%

\end{document}